\documentclass[12pt, a4paper]{article}
\usepackage[font=footnotesize]{caption}
\usepackage[utf8]{inputenc}
\usepackage{graphicx}
\usepackage{authblk}
\usepackage{booktabs}
\usepackage{verbatim}
\usepackage{url}
\usepackage{caption}
\usepackage{subcaption}
\usepackage{indentfirst}
\usepackage{url}
\usepackage{amssymb}
\usepackage{amsmath}

\usepackage{multirow}
\usepackage[table,xcdraw]{xcolor}

\usepackage{todonotes}

\usepackage{rotating}
\usepackage{array}
\title{Non-Supervised Community Detection and Hierarchical Modularity Estimation in Complex Networks}
\author{Alexandre Benatti$^1$ and \\ Luciano da F. Costa$^2$}

\affil{
$^1$Institute of Mathematics and Statistics - DCC \\
University of S\~ao Paulo \\
Rua do Mat\~ao, 1010, \\ S\~ao Paulo, SP 05508-090 Brazil %\\ \vspace{0.5cm}
\\ \vspace{0.5cm}
$^2$S\~ao Carlos Institute of Physics - DFCM \\
University of S\~ao Paulo \\
Av. Trabalhador S\~ao-Carlense, 400, \\ S\~ao Carlos, SP 13566-590 Brazil \\
(Prof. Senior)
}

\date{\emph{27th May, 2026}}

\begin{document}

\maketitle

\begin{abstract}
This work extends to complex networks a recently described methodology (A. Benatti and L. da F Costa, Detecting Hierarchical Clusters and Estimating their Modularity Directly from Dendrograms, May 2026) for non-supervised hierarchical cluster detection and hierarchical modularity estimation. First, the edge betweenness centrality of a given complex network (or graph) is estimated, and a dendrogram is obtained from these values by using some linkage criterion (average linkage is considered in the present work). The mentioned concepts and methods can then be applied to the obtained dendrogram associated with the hierarchical structure of the nodes interrelationship, paving the way to community detection and hierarchical modularity estimation. Promising results are presented and discussed respectively to varying types of modular networks, namely fractal networks (which are intrinsically hierarchical) and prime partition networks, as well as to modular networks presenting just one (or a few) hierarchical levels. 
\end{abstract}

\section{Introduction}\label{sec:introduction}

Graphs (e.g.~\cite{west2001,bollobas1998}) and complex networks (e.g.~\cite{barabasi2013network,newman2018networks,costa2011}) have been used to represent, analyze, and model several discrete theoretical and real-world structures and systems~\cite{costa2011}. Complex networks can be understood as graphs with particularly intricate properties, including hierarchical modularity. Basically, a module corresponds to a group of nodes which are more intensely interconnected with each other than with nodes in the remainder of the network. 

Among the possible operations on complex networks and graphs, the identification of their eventual \emph{modules} or \emph{communities} (e.g.~\cite{girvan2002,newman2004}) has been frequently performed in order to better understand their respective topology. This task involves intrinsic difficulties, including: (a) the communities may occur at several hierarchical scales; (b) communities are often of varying sizes (number of nodes) and may present distinct structure; (d) the number of communities is typically unknown, as is the respective main scale of interconnectivity; and (c) it has been particularly difficult to define an effective measurement of the modularity. Observe that issue (c) is a largely consequence of (a--c).

Non-supervised pattern recognition, alternatively known as \emph{clustering} (e.g.~\cite{duda2000pattern, theodoridis2006pattern}), represents another challenging area. Basically, given a set of data elements characterized by respective features, clusters correspond to groups of elements that are closer (or more similar) to each other than with the remainder of the data set.

An analogy exists between cluster detection and community finding, which can be exemplified by considering the distances between the features of data elements as links of a respective network. Therefore, the tasks of clustering and community finding become intrinsically inter-related. Actually, a matrix representing the distances (or dissimilarities) between the elements in a data set is directly analogous to a weight matrix of a complex network (assuming that the weights are associated to the dissimilarity between each pair of nodes).

Recently, a non-supervised approach~\cite{benatti2026} in which equalized merging densities of dendrograms has been described, which allows the identification of possible hierarchical clusters in a dendrogram representing a data set. Several types of modularity that can be obtained from a dendrogram were also described in that work. Due to the intrinsic relationship between clustering and community finding, it becomes possible to extend the concepts and methods described in~\cite{benatti2026} to complex networks, which provides the main motivation for the present work.

The adopted approach is based on first obtaining a weight matrix from the edge betweenness centrality of the edges of a given complex network (which can also be originally weighted). The edge betweenness centrality is assumed to be related to the topological distance (or dissimilarity) between pairs of nodes in the network. Other approaches can eventually be applied in order to map a given complex network into a respective dendrogram.

A respective dendrogram is then obtained by applying a linkage criterion. For simplicity's sake, only the average linkage approach is considered in this work. The concepts and methods described in~\cite{benatti2026} can then be applied to the dendrogram that represents the betweenness distance between the nodes in the complex network, paving the way for respective community detection and estimation of modularity.

The potential of the approach is illustrated for two types of hierarchically modular networks, namely: (a) fractal networks~\cite{benatti2024}; and (b) prime partition networks~\cite{benatti2025}. Promising results have been obtained in both cases, allowing for not only the estimation of the hierarchical clusters but also the respective overall, as well as infinitesimal and group modularities. In order to illustrate the application of the described approach also to modular networks characterized by one (or a few) hierarchical levels, it has been applied to the comparison of the overall modularity of networks composed of two interconnected modules of the types Erd\H{o}s-R\'enyi or Barab\'asi-Albert (BA, e.g.~\cite{barabasi1999BA}).

The present work starts by reviewing basic concepts from network science that are directly related to the described developments, and then revises the basic approach identification of main hierarchical levels, respective clusters, as well as the concept of infinitesimal modularity and several types of modularity obtained from it. The potential of the approach is then illustrated respectively to a type of fractal network, a prime partition network, as well as the comparison of the hierarchical structure and modularity of networks composed of two modules of the Erd\H{o}s-R\'enyi and Barab\'asi-Albert types. The work concludes with a review of the main results obtained and some perspectives for further developments.

\section{Basic Related Concepts}

\emph{Graphs} (e.g.~\cite{west2001}) are discrete structures composed of $N$ \emph{nodes} interconnected (directionally or not) by $E$ \emph{links} or \emph{edges}. The number of links emanating from a node is said to be its \emph{out-degree}, and the number of links converging at a node corresponds to its \emph{in-degree}. In non-directed graphs, the in-degree is equal to the out-degree, being called simply \emph{degree}. Weights can be associated to the links of a graph, yielding a \emph{weighted graph}.

\emph{Complex networks} (e.g.~\cite{barabasi2013network,newman2018networks,costa2011}) can be understood in an intuitive manner, as graphs presenting a particularly intricate topology respectively to a regular or uniformly random graph. Graphs characterized by hierarchical modularity can therefore be understood as complex networks.

The interconnections of a complex network can be represented in several manners, including its \emph{adjacency matrix} (especially in the case of non-weighted networks), as well as a weight matrix (in the case of weighted networks).

The \emph{betweenness centrality}~\cite{freeman1978,girvan2002} of an edge $e$ in a complex network corresponds to a measurement related to the number of minimum paths that cross that edge. Edge betweenness centrality has been related to the modularity of complex networks, especially in algorithms for community detection, including Girvan-Newman (e.g.~\cite{girvan2002}), which involves the successively removal of links with higher betweenness centrality (which tend to correspond to bridges between the main communities).

The edge betweenness centrality can be expressed as:
\begin{align}
B(e) = \sum_{i \neq j} \frac{n_{i,j}(e)}{n_{i,j}},
\end{align}
where $e$ is the specific edge in the network, $i$ and $j$ are the source and target nodes, respectively, of a path in the network, $n_{i,j}$ is the total number of shortest paths between $i$ and $j$, and $n{i,j}(e)$ is the number of shortest paths between $i$ and $j$ that pass through $e$.

The described approach involves calculating the edge betweenness centrality only once for each given network, so that the weight of each edge corresponds to the respective betweenness centrality, which is understood as a dissimilarity or, informally, as a kind of `distance' since it does not satisfy all the properties required for mathematical distance. This approach can be used for both directed and non-directed networks. However, given that the calculation of the betweenness for large structures can require a relatively high computational cost, it is possible to consider, during its calculation, only the pairs of nodes that are at a maximum topological distance from each reference edge. However, this will imply a limitation to the possibly observed scales of node interrelationship (limiting the extent of the variable $s$).

Given a set of data elements characterized by respective measurements, or features, it is possible to apply an agglomerative approach based on some linkage criterion in order to obtain a respective \emph{dendrogram} (e.g.~\cite{hill1980,duda2000pattern,theodoridis2006pattern,de2013pattern}). Possible approaches include but are not limited to single, complete, and average linkage (e.g.~\cite{duda2000pattern,theodoridis2006pattern}), which can lead to distinct but related dendrograms. The latter approach (average linkage) is adopted henceforth in this work. In addition, the topology of these dendrograms is assumed to correspond to a binary tree.

The dendrograms considered in this work have their vertical axis associated with a scale variable $s \in [0,1]$. The higher values of $s$ are associated with larger scales of interrelationship between the network nodes. The original data elements are represented, in the lowest hierarchy, as the \emph{leaves} of the respective dendrogram.

Therefore, edges with larger betweenness centrality are considered more distant (or dissimilar), allowing to take into account the bottlenecks in the original network. In particular, the largest dissimilarity has been normalized to the value $1$. A respective dendrogram is then obtained by applying the average linkage criterion to the dissimilarity matrix above.

Other types of transformations from networks to dendrogram can be considered, including mapping the distances (e.g.~topological) or similarities between the nodes of the given network.

\section{Cluster Detection and Modularity Estimation from Dendrograms}

Given a dendrogram along a dissimilarity scale $s \in [0,1]$, the values of this variable where branches occur can be identified, each being represented as a respective unit Dirac delta along a resulting function $f(s)$. Weights are then assigned to each of these Dirac delta pulses, which can correspond to the number of leaves covered by each respective branch (it is also possible to consider geometric/harmonic averages or minimum/maximum between the number of leaves in those branches). The use of the number of covered leaves as the weighting criterion is henceforth adopted in this work. This operation results in a new function $q(s)$ corresponding to a sum of Dirac delta functions weighted as described above.

By convolving the function $q(s)$ with a Gaussian with standard deviation $\tilde{\sigma}$, a smoothed and interpolated \emph{merging density function} $p(s)$ can be obtained representing $q(s)$ within a resolution $\tilde{\sigma}$. Other types of smoothing, especially methods based on non-linear diffusion (e.g.~\cite{perona1994,black1998,weickert1998}) can also be considered, allowing the peaks along $p(s)$ to be better preserved along the smoothing procedure.

The \emph{main merging levels} of the dendrogram are estimated by detecting the maximum peaks along $p(s)$. Each of these main levels is flanked by two delimiting levels, which correspond either to the extremities of the variable $s$ or the minimum peaks of $p(s)$. Associated intervals are thus defined along $s$, each corresponding to an associated main merging level. The leaves covered by each of the branches cut by the upper delimiting level associated to an interval can then be identified as corresponding to respective clusters at that respective scale of interrelationship.

The described approach performs the identification of the main merging levels from $p(s)$, simultaneously considering all scales $s \in [0,1]$. This allows the interconnectivity at each main scale to be accommodated into the analysis, avoiding interferences between scales as could happen in case the search for the main levels were to be performed by interactively partitioning the elements at a fixed scale, while trying to optimize the modularity. It could happen, for instance, that the modules at one of the main levels would interact with those at another level, possibly leading to secondary local maxima during modularity optimization.

As described in~\cite{benatti2026}, an additional approach to detecting the main merging levels can be applied which considers the signature of infinitesimal modularity of the given dendrogram.

The concepts and methods in~\cite{benatti2026} can also be used to estimate several types of modularity of the given dendrogram, including infinitesimal, average, overall, and group modularities. The \emph{infinitesimal modularity} reflects the balance and size of the groups at a specific topological scale $s$, but not the extension of the branches along the scale variable.

The extent of the stems associated to clusters can then be quantified by integrating the infinitesimal modularity along the scale variable $s$. The \emph{average modularity} consists of taking average of the infinitesimal modularity along a given scale interval $[a,b]$. In case $a=0$ and $b=1$, the average modularity becomes the \emph{overall modularity}, which provides an indication about the general modular complexity of a dendrogram. The concept of \emph{individual modularity} aims at quantifying the modularity of any given group (stem) in the dendrogram in terms of the integration of its infinitesimal modularity taken along the extent of the respective stem in the dendrogram. The extension of the individual modularity to more than one group has been addressed in terms of the \emph{set modularity}, while the \emph{groups modularity} then corresponds to the set modularity when all groups identified by a given slice of the dendrogram are taken into account.

\section{Results and Discussion}

This section illustrates the potential of the described approach to community identification and hierarchical modularity estimation with respect to two types of structures: fractal networks and prime partition networks.

\subsection{Fractal Networks}

Fractal networks have been the subject of related approaches including, but not limited to~\cite{song2005self,barabasi2001deterministic,dorogovtsev2002pseudofractal}. In particular, an approach has been described~\cite{benatti2024} in which fractal networks are obtained by recursively incorporating a generating motif with one of the nodes considered as reference into the growing structure. This type of network tends to present modules along a hierarchical scale of interconnections.

\begin{figure}[h]
  \centering
     \includegraphics[width=0.9 \textwidth]{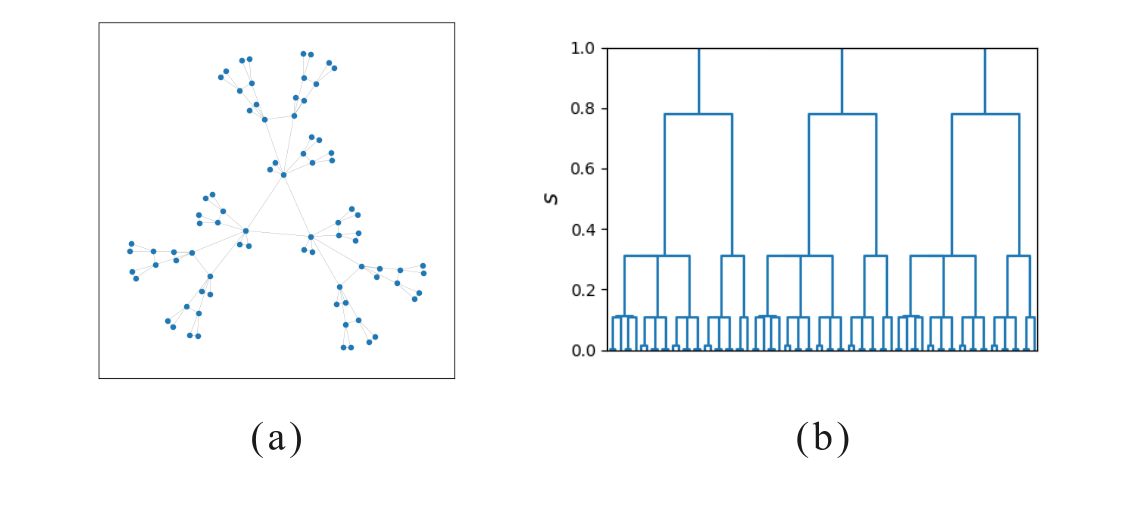}
 \caption{A fractal network (a) and an associated dendrogram visualized by the Fruchterman and Reingold methodology (e.g.~\cite{fruchterman1991}), and the dendrogram (b) obtained as described in the text.}\label{fig:fractal_net_1}
\end{figure}

Figure~\ref{fig:fractal_net_1}(a) presents a fractal network obtained for a triangular motif applying 3 interactions, as well as an associated dendrogram (b) obtained for $\tilde{\sigma} = 0.05$.

Due to the `crystalline' structure of the considered fractal networks, the branches at each main merging level can be observed to take place at the same value of the scale variable $s$. Therefore, their respective dispersions are zero.

\begin{figure}[h]
  \centering
   \includegraphics[width=0.9 \textwidth]{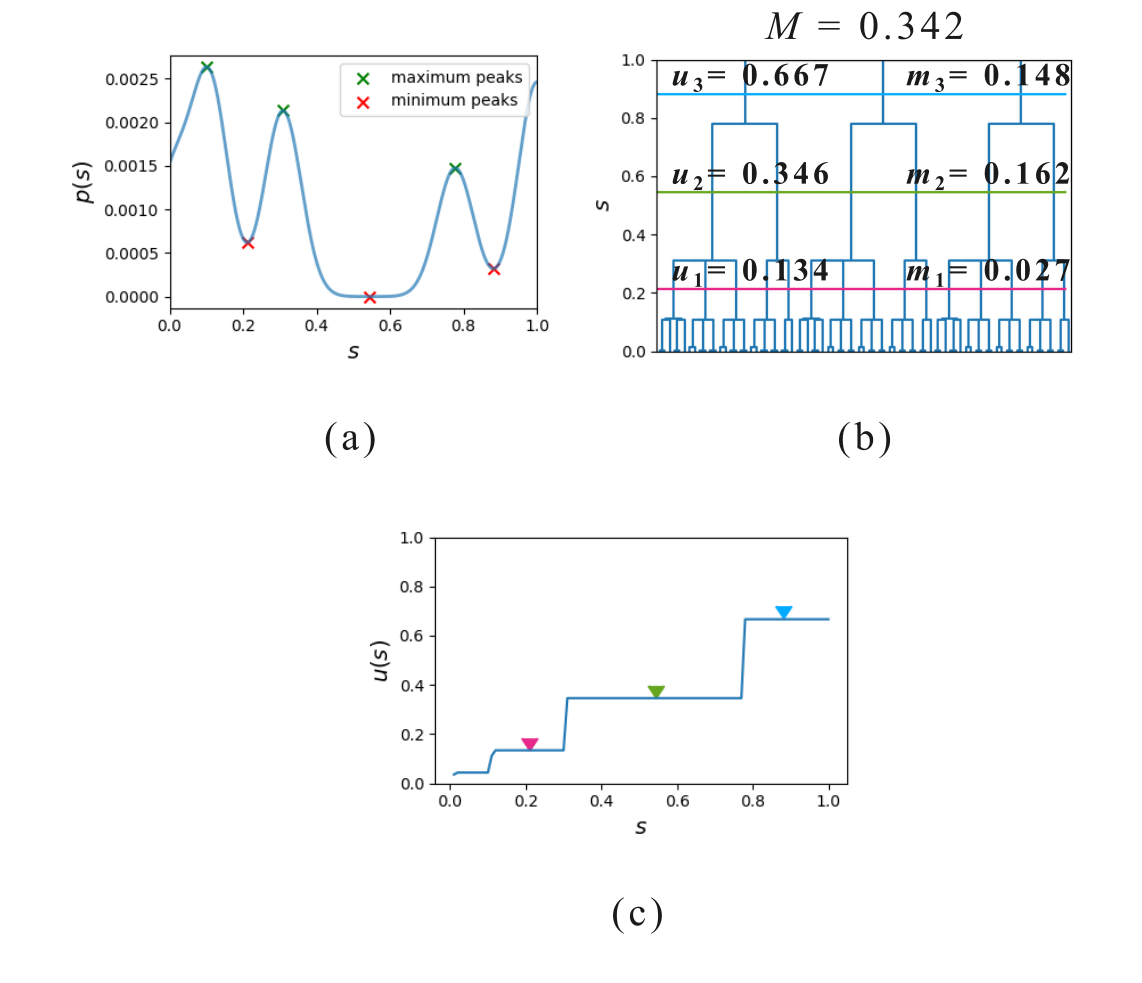}
 \caption{The balanced merging density function $p(s)$ (a) obtained for the dendrogram in Figure~\ref{fig:fractal_net_1}(b), as well as the delimiting levels overlaid onto the original dendrogram (b). The infinitesimal and groups modularities are also shown in (b). The signature of infinitesimal modularity obtained for the same considered dendrogram is shown in (c), together with marks indicating the obtained main merging levels.}\label{fig:fractal_net_2}
\end{figure}

The application of the considered methodology, adopting the weighting by sum and $\tilde{\sigma}=0.05$, allowed the estimation of the balanced merging density function shown in Figure~\ref{fig:fractal_net_2}, along which its maximum and minimum peaks are identified, resulting in the respective \emph{main merging levels} and \emph{the delimiting levels}, the latter shown in (b) overlaid onto the original dendrogram. The overall ($M$) as well as the infinitesimal ($u$) and group ($m$) modularities estimated for each of the detected main merging levels are also shown in the figure.

\begin{figure}[h]
  \centering
     \includegraphics[width=1 \textwidth]{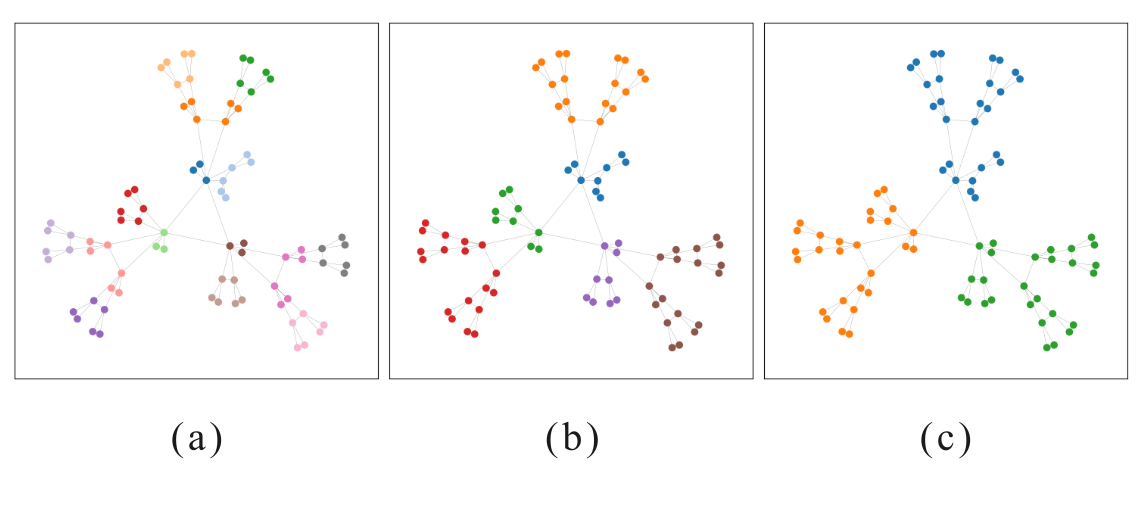}
 \caption{The clusters identified respectively to the three successive detected main merging scales.}\label{fig:fractal_net_3}
\end{figure}

The clusters detected at each of the three main merging scales are shown in Figure~\ref{fig:fractal_net_3}, which can be observed to correspond to groups of neighboring nodes in the original network shown in Figure~\ref{fig:fractal_net_1}(a). Progressive merging of the subgroups that takes place along the three succeeding main scales can be observed.

Regarding the obtained modularities, as shown in Figure~\ref{fig:fractal_net_3}(b), a moderate overall modularity of $M\approx 0.342$ can be observed. These results can be better understood by referring to Figure~\ref{fig:fractal_net_3}(c), which shows the infinitesimal modularities obtained for the same dendrogram. Despite the almost maximum balance of the number of elements in all obtained groups, the infinitesimal modularity is smaller than $1$ because of the average number of groups at each hierarchy being smaller than 2. The obtained groups modularity indicates that the groups at the second main hierarchical level resulted with the highest group modularity, which is a consequence of their stems extending along a longer interval along the scale variable $s$.

\section{Prime Partition Networks}

In this section, the concepts and methods described in the present work for community detection and hierarchical modularity estimation are illustrated respectively to prime partition networks.

The \emph{Goldbach conjecture} (e.g.~\cite{wang2002}) postulates that it is possible to express any integer number as the sum of two prime numbers, which corresponds to a \emph{partition} of that integer number. The \emph{weak Goldbach conjecture} (e.g.~\cite{wikipedia_goldbach_weak}) proposes that any odd integer number can be obtained from the sum of three prime numbers. More recently~\cite{benatti2025}, it has been postulated that two of the three prime numbers involved in the weak Goldbach conjecture can always be identical, leading to a \emph{generalized Goldbach conjecture} suggesting that any integer value can be expressed as the sum of two prime numbers or as the sum of a prime number and twice a prime number (double primes).

\begin{figure}[h]
  \centering
     \includegraphics[width=0.4 \textwidth]{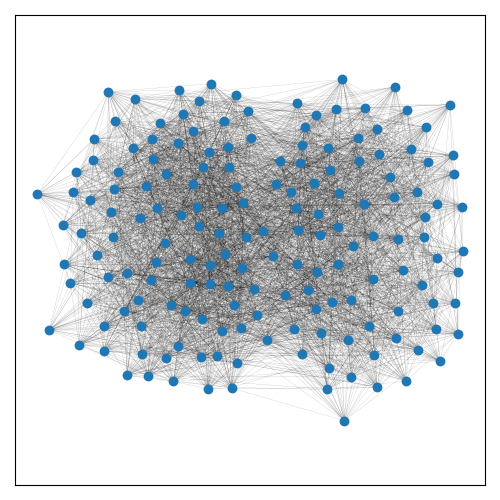}
 \caption{The prime partition network up to the prime number $P=997$ visualized by using the Fruchterman and Reingold methodology (e.g.~\cite{fruchterman1991}) considering non-directed links.}\label{fig:prime_net_1}
\end{figure}

The above mentioned generalized Goldbach conjecture has been used to obtain graph representations of relationship between prime numbers, called \emph{prime partition networks}, in which each prime number up to a limiting value $P$ is represented as a node, and directed edges are established from the prime and double prime values toward the prime numbers to which they can contribute as terms of the sums in the respective partitions. Figure~\ref{fig:prime_net_1} illustrates a prime partition network for the prime numbers from 2 to 997, visualized as a non-directed graph.

As discussed in~\cite{benatti2025}, prime number networks present a strong hierarchical modular organization. Although the main hierarchical level could be visually identified, a better understanding of the topology of these networks can be obtained by applying methods for identifying their hierarchical structure and estimating their hierarchical modularity. This has been addressed as follows.

\begin{figure}[h]
  \centering
   \includegraphics[width=0.99 \textwidth]{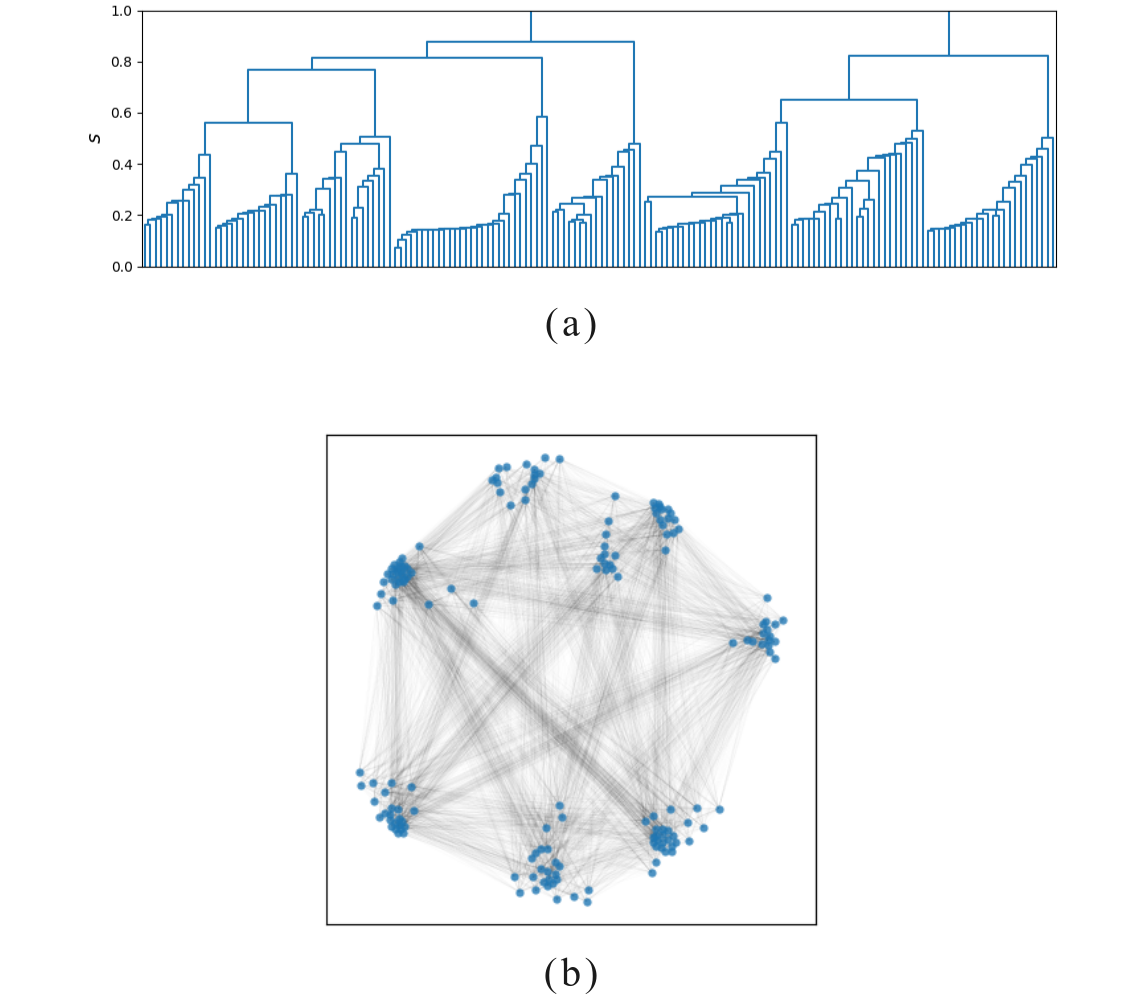}
 \caption{The dendrogram obtained for the considered prime partition network (a) and the visualization of the respective network considering the position of the nodes determined by cophenetic distance and multidimensional scaling.}\label{fig:prime_net_2}
\end{figure}

Figure~\ref{fig:prime_net_2}(a) presents the dendrogram obtained for the prime partition network in Figure~\ref{fig:prime_net_1} by using the adopted methodology, which considers the edge betweenness between the original network nodes. This dendrogram resulted strongly modular, while also presenting a particularly complex modular structure along the scale variable $s$. The visualization of the prime partition network considering its original interconnections but with positions obtained by using multidimensional scaling (e.g.~\cite{kruskal1964}) of the symmetric weight matrix while using the cophenetic distance (see~\cite{benatti2026,sokal1962}) from the betweenness values obtained considering the directionality of the original network is shown in (b).

Interestingly, the visualization obtained indicates strong modularity and well-separated communities, which is in agreement with the overall structure of the respective dendrogram. The edge betweenness centrality adopted in the analysis enhanced the original modularity of the original directed network, allowing the identification of the respective communities.

\begin{figure}[h]
  \centering
   \includegraphics[width=0.99 \textwidth]{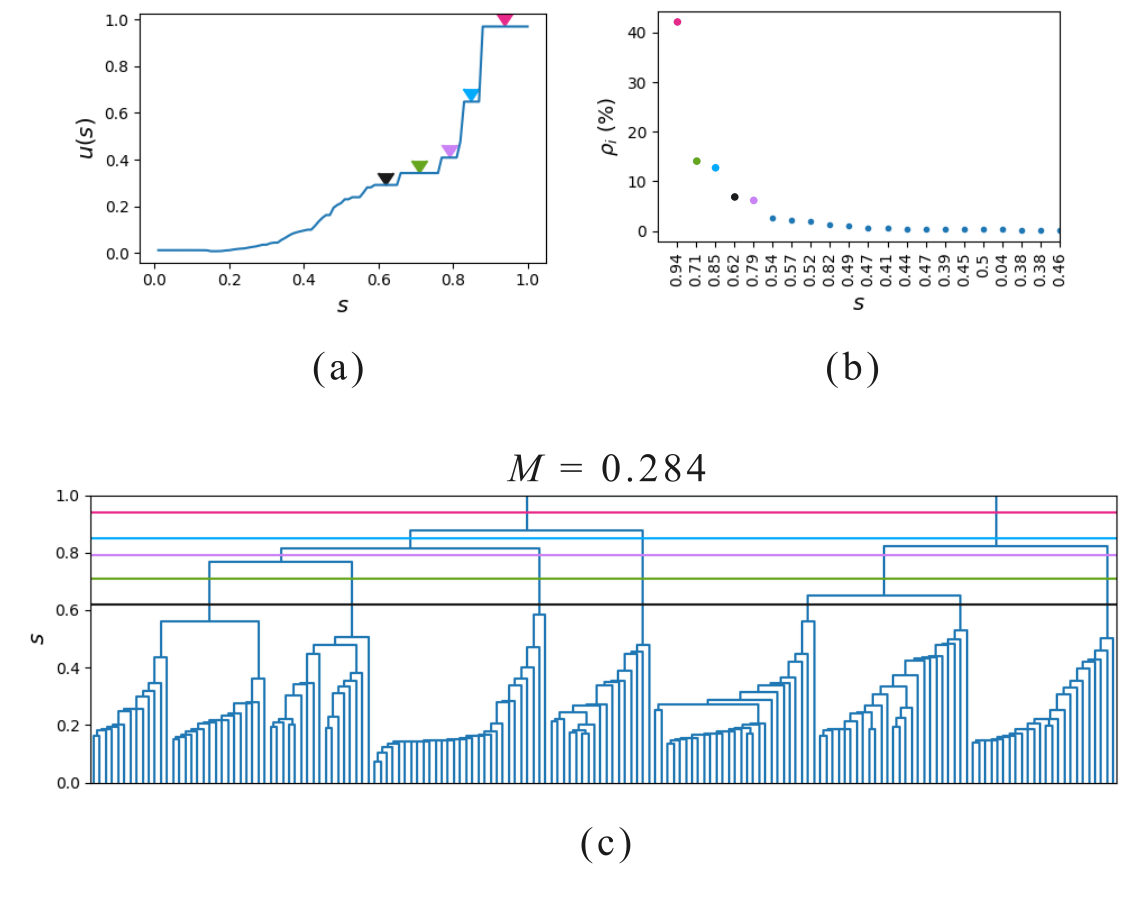}
 \caption{The signature of individual modularity obtained for the prime partition network is shown in (a), while their sorted relevance values are presented in (b). The first five most relevant merging levels are identified in (c).}\label{fig:prime_net_3}
\end{figure}

The infinitesimal modularity signature obtained for the respective dendrogram is also presented in Figure~\ref{fig:prime_net_3}(a). The sorted relevance of the merging levels is shown in (b). In addition to the infinitesimal modularity, the relevance index $\rho$ also takes into account the length of the stems along the scale variable, so that the infinitesimal modularity of a set of groups at a given level modularity may not correlate with the relevance of these clusters.

Interestingly, the relevance of the merging levels decreases steadily, therefore confirming the strong modularity of the prime partition networks. A total of 5 particularly relevant merging levels can be identified, corresponding to the first values along the distribution shown in Figure~\ref{fig:prime_net_3}(b). The most relevant merging level obtained, which has the longest stem and combined sizes of groups, corresponds to the bilateral symmetry previously identified in Figure~\ref{fig:prime_net_1} for prime partition networks. 

\begin{figure}[h]
  \centering
   \includegraphics[width=0.99 \textwidth]{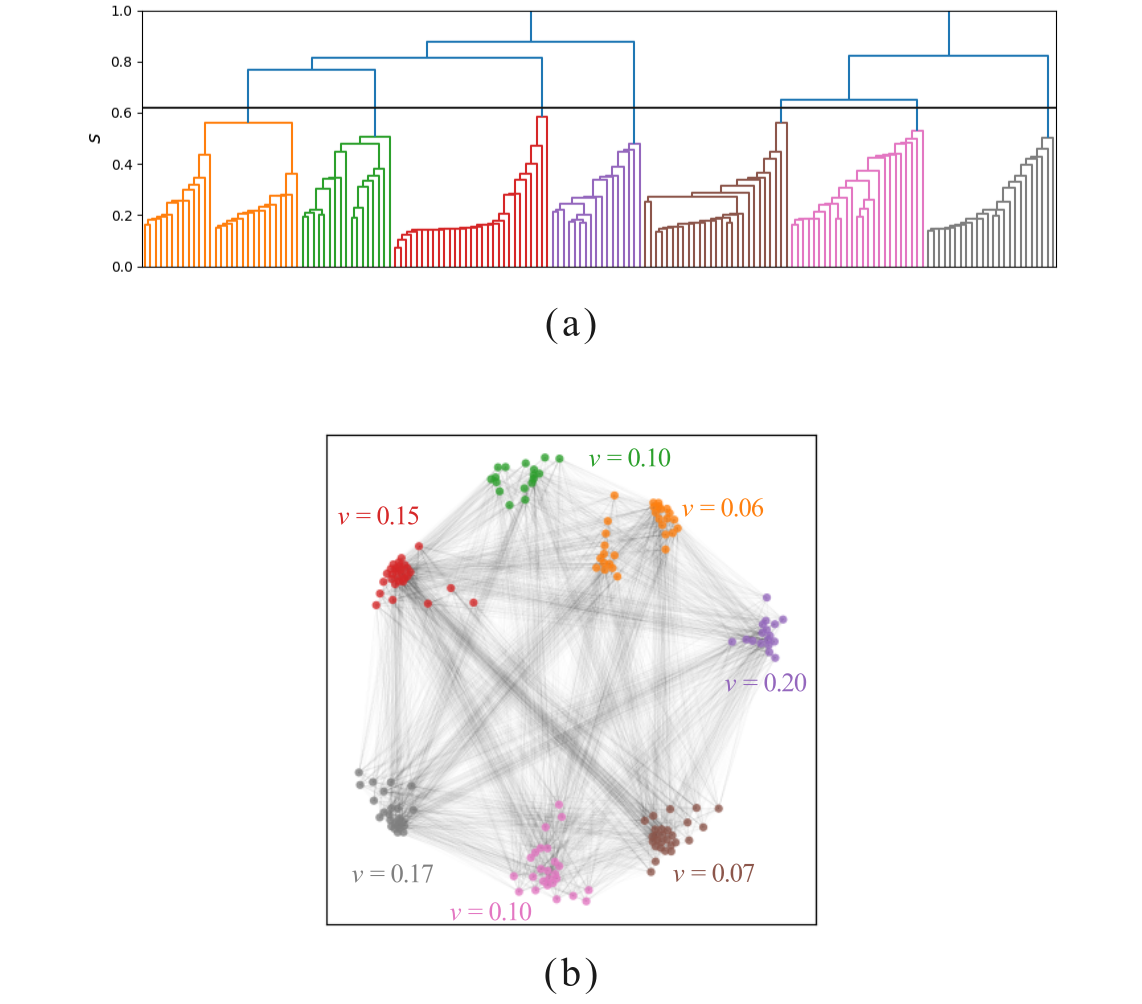}
 \caption{Dendrogram of the prime partition network showing the slice respective to $s\approx 0.618$ (a) and the respectively identified communities (b).}\label{fig:prime_net_4}
\end{figure}

For each of the identified most relevant merging levels, the respective communities can be obtained as the set of nodes in each of the sliced stems.
The dendrogram and nodes partition obtained for the slice of the dendrogram at $s \approx 0.618$, which is the fourth most relevant merging level, are shown in Figure~\ref{fig:prime_net_4}(a). Each identified community is shown in a different color in (b), jointly with the value of its respective individual modularity. Variation of individual modularity can be observed, ranging from $\nu = 0.06$ up to $\nu = 0.20$.

\section{Comparing Modularity of Networks with ER and BA Modules}

Although the concepts and methods described in this work mostly relate to \emph{hierarchical} modular networks, they also apply to complex networks with a single (or a few) hierarchical levels. This section illustrates this possibility respectively to the comparison of the modular properties presented by a pair of interconnected modules of the types Erd\H{o}s-R\'enyi (ER, e.g.~\cite{erdos1959random}) or Barab\'asi-Albert (BA, e.g.~\cite{barabasi1999BA}) structures. Each of these two modules have the same average degree of $\left< k \right> = 17.8$, being interconnected by $e$ uniformly random undirected edges avoiding those already existing. For each of the configurations considered, the described methodology is applied in order to estimate the respective overall modularity.

\begin{figure}[h]
  \centering
     \includegraphics[width=1 \textwidth]{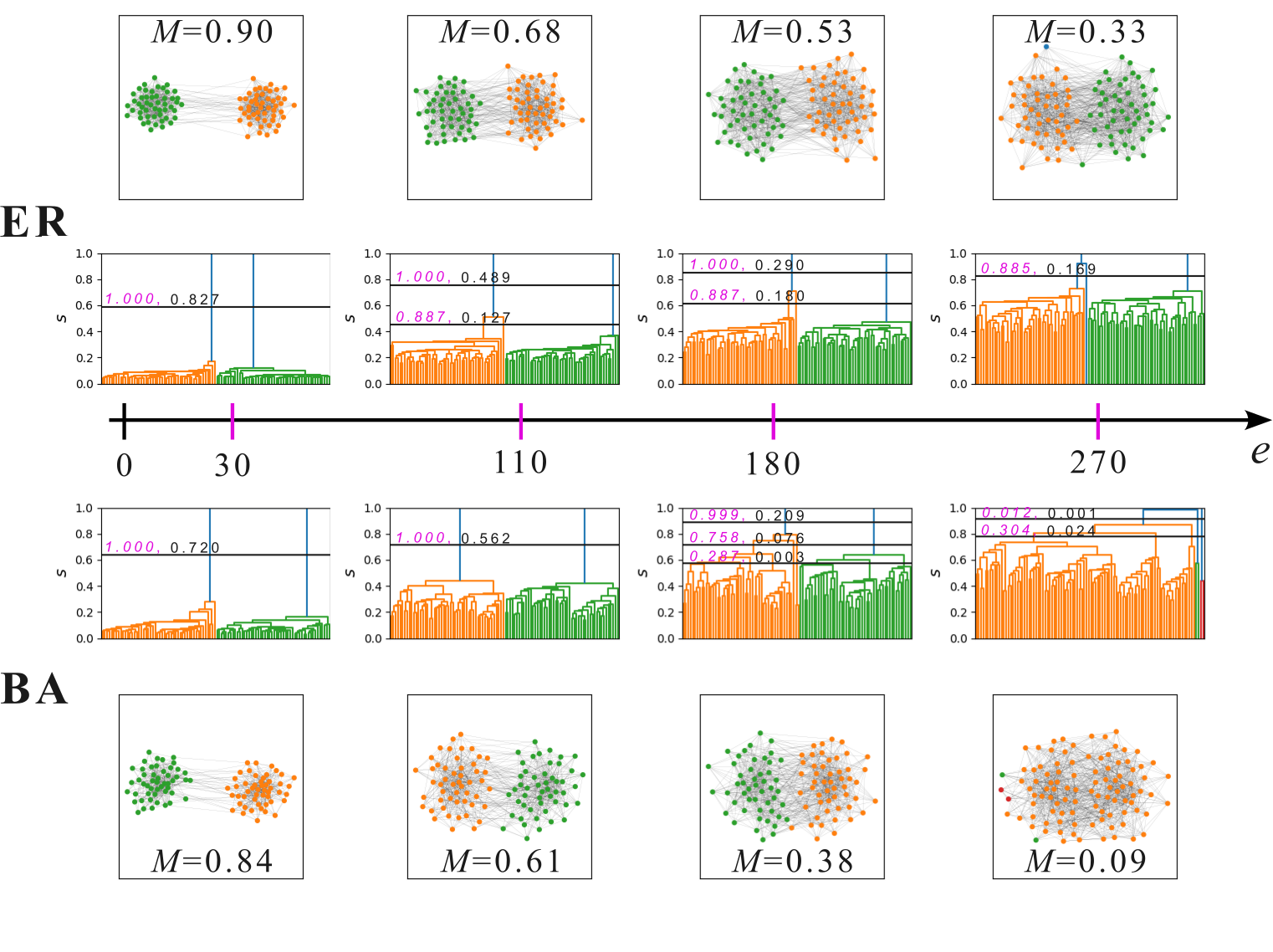}
 \caption{Illustration of the comparison of the overall ($M$), infinitesimal ($u$), and group ($m$) modularities of networks composed of two interconnected modules of types ER and BA for $e \in [30,110,180,270]$. The infinitesimal and group modularities are shown respectively to the main detected groups are shown in magenta and black, respectively. The colors of the network nodes refer to the detected modules.} \label{fig:ER_BA}
\end{figure}

Examples of the considered modular networks are illustrated in Figure~\ref{fig:ER_BA}, which shows the networks together with their respective dendrograms, with $e \in [30,110,180,270]$. Each of the modules has 50 nodes.

Although in both cases the overall modularity tends to decrease with $e$, this decrease can be observed to be faster for BA structures than ER. The results obtained also indicate that the infinitesimal and group modularity tend to be unrelated in most cases. That is a consequence of the latter type of modularity considering the common extension of the groups along the scale variable. For example, in the case of the lower hierarchical level in the ER case along the second column in Figure~\ref{fig:ER_BA}, although the three detected groups have a considerably large infinitesimal modularity, their group modularity is substantially small due to the short common length of the respective branches.

\begin{figure}[h]
  \centering
     \includegraphics[width=0.55 \textwidth]{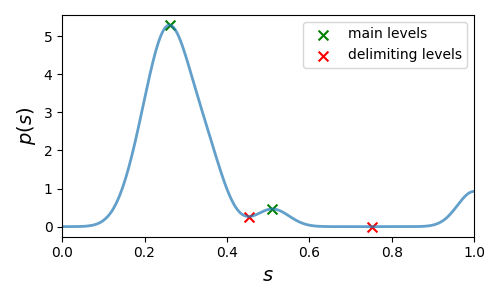}
 \caption{The equalized merging density $p(s)$ for the dendrogram obtained for the network with two ER modules for $e=110$ shown in Figure~\ref{fig:ER_BA} with.}\label{fig:den_peaks}
\end{figure}

It is interesting to observe that the situation discussed above, namely the detection of three groups, was induced during the dendrogram construction due to the presence of a single data element joining 49 other data elements. The importance of this single point is relative to the specific situation and application underlying the original data elements, in the sense that it may correspond to an important distinct data element or to a noisy or incorrect data element. Interestingly, the methodology presented in the present work provides some means for identifying these situations in which a substantially smaller group merges with a larger group. More specifically, the function $p(s)$ provides some information about the relative relevance of each subcluster. As an example, Figure~\ref{fig:den_peaks} presents the equalized merging density $p(s)$ for the dendrogram obtained for the network with ER modules for $c=110$. The peak implied by the junction of the 49 data elements and the single element is characterized by the value of $\approx 0.46$, which is much smaller than the value of $\approx 5.39$ obtained for the two detected groups.

\begin{figure}[h]
  \centering
     \includegraphics[width=.6 \textwidth]{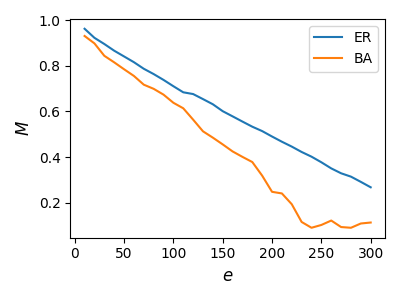}
 \caption{Overall modularity in function of $e$ for ER and BA network models.} \label{fig:Mxe}
\end{figure}

Figure~\ref{fig:Mxe} presents the average of the overall modularity obtained for the networks involving ER and BA modules in terms of $e$. The presented modularity signatures were obtained for a single sample of each type of network.

Some interesting tendencies can be observed in the results obtained. First, we find that network with BA modules tended to lose their overall modular structure more quickly as $e$ increases than the ER cases. In addition, for the considered scale of $s$, the network with BA modules reached the smallest overall modularity (near 0) sooner than the network with the ER modules.

\section{Concluding Remarks}

The search for communities and estimation of their modularity constitute two central tasks in network science. The present work described an approach to detecting communities which is based on a recent work~\cite{benatti2026} aimed at non-supervised clustering of data elements in pattern recognition directly from dendrograms. In addition, the hierarchical modularity of the dendrograms representing complex networks has also been addressed by using the concepts and methods in that work.

Before these tasks can be performed, it is necessary first to transform the given complex network into a suitable dendrogram. Though alternative approaches could be considered, here the edge betweenness centrality is taken into account in order to obtain a measurement of the `distances' (dissimilarity) between respective pairs of points, yielding a dissimilarity matrix from which a respective dendrogram associated with the given network can be estimated.

In order to exercise the ability of the considered approach to cope with hierarchical structures, two types of hierarchically modular complex networks have been considered, namely fractal networks~\cite{benatti2024} and prime number networks~\cite{benatti2025}. Promising results have been obtained in both cases. More specifically, the application of the considered methodology to fractal networks and prime partition networks allowed not only the identification of hierarchical modules, but also the estimation of infinitesimal overall and group modularities. In the case of the prime partition network, the applied approach confirmed the strong modularity and complexity previously observed~\cite{benatti2025} for this type of networks. In particular, several hierarchical levels have been identified, and the communities obtained at the most relevant of these levels have been found to have varying sizes, dispersions, and interconnectivity. These results indicated that, from the perspective of their pairwise partitions, prime numbers can be classified into groups which have distinct properties.

The described approach has also been illustrated with respect to modular networks having just one (or a few) hierarchical levels. More specifically, modular networks composed of two interconnected modules of types ER and BA have been characterized in terms of their overall modularity for successive levels of interconnectivity between the two modules. The results indicated that modular network composed of BA modules tended to have their modules less preserved as the interconnectivity between them increased.

Because the reported results are respective to specific types of networks and parametric configurations, it is necessary to further investigate the performance of the described approach respectively to more general situations, including other sizes and types of networks, parametric configuration, normalizations, among other possibilities.

Additional possibilities for further research include, but are not limited to, the consideration of non-linear diffusion to obtain $p(s)$, applying the described approach recursively (through successive zooming) in order to study given networks along successive resolutions, and considering other ways for transforming complex networks into dendrograms. In particular, the highly complex modular structure and properties observed for the prime partition networks motivate several types of further analysis.

\section*{Acknowledgments}
A. Benatti is grateful to MCTI PPI-SOFTEX (TIC 13 DOU 01245.0102\\22/2022-44), FAPESP (grant 2025/26083-7 and 2022/15304-4). Luciano da F. Costa thanks CNPq (grant no.~313505/2023-3) and FAPESP (grant 2022/15304-4).

\bibliography{ref}
\bibliographystyle{unsrt}

\end{document}